
\magnification=\magstep1
\mathchardef\sigma="011B
\def\bmit{\sigma\kern -0.6em}
\hsize=6.5truein
\vsize=8.9truein
\font\bigbf=cmbx10 scaled\magstep1
\abovedisplayskip=15pt plus 3pt minus 3pt
\abovedisplayshortskip=10pt plus 3pt
\belowdisplayskip=15pt plus 3pt minus 3pt
\belowdisplayshortskip=10pt plus 3pt minus 2pt
\mathsurround=3pt
\parindent=0pt
\baselineskip=12pt
\parskip=12pt

\def\sectionbreak{\penalty -500}
\def\section#1.#2.{\sectionbreak\vskip0.8truecm\centerline{\bf #1.\ \
#2}
                    \par\nobreak\vskip3truept}
\def\vvs{\noalign{\vskip 3pt}}
\looseness=2

\centerline{\bigbf CLOCK TIME AND ENTROPY}

\bigskip\medskip
\centerline{Don N. Page}
\centerline{CIAR Cosmology Program}
\centerline{Institute for Theoretical Physics}
\centerline{Department of Physics}
\centerline{University of Alberta}
\centerline{Edmonton, Alberta}
\centerline{Canada T6G 2J1}
\bigskip
\centerline{To be published in
 {\it Physical Origins of Time Asymmetry,} Eds. J. J. Halliwell,}
\centerline{J. Perez-Mercader, and W. H. \.Zurek, Cambridge
University Press, Cambridge (1993) }
\vskip0.5truecm
\bigskip

\section 1.What is fundamental in a quantum description of the
world?.

\hangindent=25pt\hangafter=-1 It is a matter of debate what
the basic entities of the universe
are in quantum theory. Expanding the metaphor used by Griffiths at
this conference, one might imagine that the conversations of the
three meals of the day could be restricted to three possible answers:

\hangindent=25pt\hangafter=-1 (a) Amplitudes (subject of lunch
conversation).
One might imagine formulating quantum theory in the morning and
concluding at
lunch that complex amplitudes were the fundamental entities. Each
fine-grained history $h_i$ might be assigned an amplitude
$$
A[h_i]=e^{iS[h_i]},\eqno(1)
$$
where $S[h_i]$ is the action (in units of $\hbar)$ of that history
(Feynman and Hibbs 1965). The fine-grained histories could then be
combined into a weighted group of histories $\alpha$ (called a
coarse-grained history) with amplitude
$$
A[\alpha]=\sum_i w_\alpha\,[h_i] e^{iS[h_i]},\eqno(2)
$$
where $w_\alpha$ is an $\alpha\hbox{-dependent}$ weight functional of
the
fine-grained histories $h_i.$  For example, $A[\alpha]$ could be
given by
a path integral over a set of histories, so that $w_\alpha$ could be
$1$
if $h_i$ is in the set and $0$ if $h_i$ is not. More generally,
$w_\alpha[h_i]$ might be weighted by the complex amplitude for the
initial configuration of the history $h_i.$ If $\alpha$ consisted of
all
of the histories that reached a certain final configuration, and no
others, $A[\alpha]$ would be what is called the wavefunction
evaluated at
that final configuration. A problem with these amplitudes as basic
entities, however, is that it is not clear what direct interpretation
they have.

\hangindent=25pt\hangafter=-1 (b) Probabilities for decohering sets
of
histories (subject of
dinner conversation) (Griffiths 1984; Omn\`es 1988a, 1988b, 1988c,
1989; Gell-Mann and Hartle 1990; Hartle 1990a, 1990b, 1991). One
might
imagine interpreting the amplitudes in the afternoon so that the
(unnormalized) probability associated with each coarse-grained
history $\alpha$ is
$$
p(\alpha)=|A(\alpha)|^2.\eqno(3)
$$
In order to get a set of probabilities obeying the usual axioms of
probability, one then needs a set of $\alpha\hbox{'s}$ such that for
any
pair $(\alpha\ne \alpha^\prime),$
$$
p(\alpha+\alpha^\prime)
\equiv
|A(\alpha)+A(\alpha^\prime)|^2=p(\alpha)+p(\alpha^\prime)\equiv
|A(\alpha)|^2+|A(\alpha^\prime)|^2.\eqno(4)
$$
This means that the real part of the interference terms must vanish,
$$
2\,\hbox{Re}\,A(\alpha)A^*(\alpha^\prime)
=A(\alpha)A^*(\alpha^\prime)+A^*(\alpha)A(\alpha^\prime)=0.\eqno(5)
$$
More generally, one may have
a decoherence functional $D(\alpha,\alpha^\prime)$ with
$$
\hbox{Re}\,D(\alpha,\alpha^\prime)=0,\eqno(6)
$$
and then
$$
p(\alpha)=D(\alpha,\alpha).\eqno(7)
$$
One then calls this set of $\alpha\hbox{'s}$ a decohering set of
(coarse-grained) histories. One says that probabilities are defined
only for each member of such a decohering set. Then the viewpoint is
that such probabilities are the fundamental entities of the quantum
theory. Feeling that his task is basically completed, the theorist
retires for the night.

\hangindent=25pt\hangafter=-1 (c) Testable conditional probabilities
(subject
of breakfast
conversation) (Page and Wootters 1983; Page 1986a, 1986b, 1987, 1988,
1989, 1991a, 1991b, 1992a).
After a good night's sleep, the theorist, able to think more clearly,
realizes that even decohering sets of histories are too broad to be
directly interpreted. Instead, it is a much narrower set of
conditional probabilities that we, living within the universe, can
test. So far as we know, each of these occurs on a single
hypersurface (at a single ``time''), and perhaps is also highly
localized on that hypersurface.

\hangindent=25pt\hangafter=-1 That is, we cannot directly compare
things
at different times, but
only different records at the same time. We cannot know the past
except through its records in the present, so it is only present
records that we can really test. For example, we cannot directly test
the conditional probability that the electron has spin up at $t=t_f,$
given that it had spin up at $t=t_i<t_f,$ but only given that there
are records at $t=t_f$ that we interpret as indicating the electron
had spin up at $t=t_i.$ Wheeler (1978) states this even more
strongly: ``the past has no existence except as it is recorded in the
present.''

\hangindent=25pt\hangafter=-1 This principle should perhaps also be
extended to say that we cannot
directly compare things at different locations either, so that all
observations are really localized in space as well as in time.
Certainly each of my observations seems to be localized within the
spatial extent of my brain and the temporal extent of a single
conscious moment, though it appears to be an exaggeration to say it
is localized to a single spacetime point. Whether it is actually
localized on a single hypersurface of a certain spatial extent is
admittedly an open question, but since the duration of a single
conscious moment is so much shorter than the apparent age of the
universe, it seems reasonable and consistent with all we do know to
idealize each observation as occurring at a single time, on a single
spatial hypersurface.

\hangindent=25pt\hangafter=-1 If the quantum state of the universe on
a spatial
hypersurface is
given by a density matrix $\rho,$ the conditional probability of the
result $A$, given a testable condition $B,$ is
$$
p(A|B)={\hbox{Tr}\,(P_A P_B\,\rho\,P_B)\over
\hbox{Tr}\,(P_B\,\rho\,P_B)}
={\hbox{Tr}\,(P_B P_A
P_B\,\rho)\over \hbox{Tr}\,(P_B\,\rho)},\eqno(8)
$$
where $P_A=P_A^{\dagger}=P_A^2$ and $P_B=P_B^{\dagger}=P_B^2$ are the
corresponding
projection operators. The testable conditional probabilities are then
the subject of rational lunch conversation.

\hangindent=25pt\hangafter=-1 Of course, the theorist's job is not
completed
with the mere
formulation of Eq.~(8): one must discover the density matrix of the
universe, formulate the projection operators, and calculate the
quantities in Eq.~(8). Then there is the question of which results
and conditions are testable. One could avoid this problem by simply
postulating that the conditional probabilities associated with all
projection operators $P_A$ and $P_B$ are meaningful fundamental
entities. However, most of these would not be readily interpretable,
so there would seem to be little motivation not to go back to the
complex amplitudes (1) and (2) as more fundamental.

\hangindent=25pt\hangafter=-1 A conclusion of these various meal
discussions is
that amplitudes
seem to be the most basic entities, but that if one wishes a more
restricted set that can be readily interpreted and tested, it does
not seem to be going far enough to say they are the probabilities for
decohering sets of histories. Testable conditional probabilities
appear in reality to be restricted to a single hypersurface, not to
histories, though even this restriction alone is almost certainly not
enough.

\vfill\eject
\parindent=25pt
\section 2. Inaccessibility of coordinate time.
\hangindent=25pt\hangafter=-1 As mentioned above, testable
conditional
probabilities (8) seem to be
confined to a single ``time'' (e.g., a spatial hypersurface in
canonical gravity). Yet they cannot depend on the value of the
coordinate time labeling the hypersurface, which is completely
unobservable. One can give three arguments for this fact:

\hangindent=25pt\hangafter=-1
(a) For a closed universe, the Wheeler-DeWitt equation,
$H\psi=0,$ implies that $\psi$ is independent of $t.$

\hangindent=25pt\hangafter=-1
(b) For an asymptotically-flat open universe, the long-range
gravitational field provides a superselection rule for the total
energy, just as the long-range Coulomb electric field provides a
superselection rule for the total charge (Strocchi and Wightman
1974). This means that phases between states of different energy are
unmeasurable, so the coordinate-time dependence of the density matrix
$\rho$ is not detectable (Page and Wootters 1983).

\hangindent=25pt\hangafter=-1
(c) In any case, one has no access to the coordinate time
$t,$ so one should average over this inaccessible variable to get a
density matrix for the other variables (Page 1989). This
time-averaged density matrix
in the Schr\"odinger picture is
$$
\overline\rho=\langle\rho(t)
\rangle\equiv \lim_{T\to\infty} {1\over 2T}\int_{-T}^{T}
dt\,\rho(t)\eqno(9)
$$
and in the Heisenberg picture is
$$
\overline\rho=\langle e^{-iHt}\rho
e^{iHt}\rangle\equiv\lim_{T\to\infty}{1\over
2T}\int_{-T}^T dt\,e^{-iHt}\rho\,e^{iHt}.\eqno(10)
$$
Without access to coordinate time $t,$ we can only test conditional
probabilities of the form
$$
\eqalignno
{p(A|B) &={\langle\hbox{Tr}\,
[P_B P_A P_B\,\rho(t)]\rangle\over
\langle\hbox{Tr}\,(P_B\,\rho(t))\rangle}=
{\hbox{Tr}\,(P_B P_A P_B\,\overline\rho)\over
\hbox{Tr}\,(P_B\,\overline\rho)}
={\langle\hbox{Tr}\,(P_B P_A P_B
e^{-iHt}\rho(0)\,e^{iHt})\rangle\over
\langle\hbox{Tr}\,(P_B e^{-iHt}\rho(0)\,e^{iHt})\rangle}
&\cr\noalign{\vskip 3pt\medbreak}
&={\hbox{Tr}\,[\langle e^{iHt} P_B P_A
P_B\,e^{-iHt}\rangle\rho(0)]\over
\hbox{Tr}\,[\langle e^{iHt} P_B\,e^{-iHt}\rangle\rho(0)]}
={\hbox{Tr}\,[\overline{P_B P_A P_B}\,\rho(0)]
\over \hbox{Tr}\,[\overline P_B\,\rho(0)]}, &(11)\cr
}
$$
where in the Schr\"odinger picture
$$
\overline O\equiv \langle e^{iHt}O e^{-iHt}\rangle\eqno(12)
$$
for an observable $O$ without explicit time dependence,
and in the Heisenberg picture
$$
\eqalignno{\overline O\equiv \langle O(t)\rangle &=\langle e^{iHt}
O(0)
e^{-iHt}\rangle.
 &(13)\cr\vvs
 [H,\overline O] &=0, &(14)\cr}
 $$
 so the observable $\overline O$ is stationary.

\hangindent=25pt\hangafter=-1 Thus testable conditional probabilities
depend
only on the stationary
 (time-averaged) density matrix $\overline\rho,$ or alternatively,
only on
$\rho(0)$ and stationary observables like $\overline{P_B}$ and
$\overline{P_B
P_A P_B}\ \ (\ne \overline{P_B}\,\overline{P_A}\,\overline{P_B}).$ In
a
constrained
Hamiltonian system like canonical quantum gravity, these observables,
but not the individual projection operators, would commute with the
constraints (e.g., the Wheeler-DeWitt operator).

\section 3.Observable evolution as dependence on clock time.
\hangindent=25pt\hangafter=-1 Although unobservable coordinate time
cannot be
part of the condition
\ \ B\ \ of $p(A|B)$ given by Eq.~(8) or (11), the reading of a
physical clock can be. The dependence of $p(A|B)$ upon the clock
reading is then the observable time evolution of the system.

\hangindent=25pt\hangafter=-1 For simplicity, consider the case where
the
condition is entirely the
reading of a clock subsystem $(C)$ with states $|\psi_C(T)\rangle.$
Let
$$
P_B\to P_T=|\psi_C(T)\rangle \langle\psi_C(T)|\otimes I_R,\eqno(15)
$$
where $I_R$ is the unit operator for the rest $(R)$ of the closed
system (e.g., the rest of the universe). Then
$$
p(A|T)={\hbox{Tr}\,
(P_T P_A P_T\,\overline\rho)\over \hbox{Tr}\,(P_T
\,\overline\rho)}\eqno(16)
$$
can vary with the clock time $T.$

\hangindent=25pt\hangafter=-1 In the special case that the clock does
not
interact with the rest of
the system, so
$$
H=H_C\otimes I_R+I_C\otimes H_R,\eqno(17)
$$
and in the case that the result  $A$ does not concern the clock,
so
$$
P_A=I_C\otimes P_{AR}\eqno(18)
$$
with $P_{AR}$ acting only on $R,$
then the choice of clock states of reading $T$ as
$$
|\psi_C(T)\rangle\equiv e^{-iH_C T}\,|\psi_C(0)\rangle\eqno(19)
$$
leads to the following familiar-looking equation for the conditional
probability of $A$ given the clock reading $T:$
$$
p(A|T)=\mathop{\hbox{tr}}_R\,[P_{AR}\,\rho_R(T)].\eqno(20)
$$
Here, the conditional density matrix for $R$ at clock reading $T$ is
$$
\eqalignno{\rho_R(T) &=\mathop{\hbox{tr}}_C
(P_T\,\overline\rho\,P_T)/\hbox{Tr}\,(P_T\,\overline\rho)
&\cr\vvs
&=e^{-iH_R T}\rho_R(0) e^{iH_R T}, &(21)\cr}
$$
thus evolving (in the Schr\"odinger picture) according to the von
Neumann equation with respect to the clock time $T$ (Page and
Wootters 1983).

\hangindent=25pt\hangafter=-1 In the more general case that the clock
does
interact with the rest
of the system and/or if the clock states are not defined by Eq.~(19)
(which gives
$$
P_T=e^{iHT} P_{T=0} e^{-iHT}\eqno(22)
$$
and has the somewhat undesirable consequence that $P_T\,P_{T^\prime}$
is
not necessarily zero for $T\ne T^\prime),$ then generically  $p(A|T)$
can
be obtained from Eq.~(16), but it will not have the
$\hbox{tr}\,(P_{AR}\,\rho_R)$ form of Eq.~(20) with a unitarily
evolved
density matrix $\rho_R(T).$ In this more general case clock time will
not have been defined so that motion is so simple as Eqs.~(20) and
(21), in contrast to the criterion of Misner, Thorne, and Wheeler
(1973).

\hangindent=25pt\hangafter=-1 Karel Kucha\v r (1992, and at lunch
during this
conference) has given
two objections to the conditional probability interpretation outlined
here:

\hangindent=25pt\hangafter=-1 (a) The application of the condition
violates the
constraints.
Eq.~(8) or (11) may be written as
$$
p(A|B)=\hbox{Tr}\,(P_A\,\rho_B),\eqno(23)
$$
where the conditional density matrix
$$
\rho_B=P_B\,\rho\,P_B/\hbox{Tr}\,(P_B\,\rho)\eqno(24)
$$
generally does not satisfy the constraints, e.g.,
$$
[H,\rho_B]\ne 0,\eqno(25)
$$
if
$$
[H,P_B]\ne 0.\eqno(26)
$$
However, $\rho_B$ is merely a calculational device to simplify
Eq.~(8) or (11) to Eq.~(23). One could instead rewrite Eq.~(8) or
(11) as
$$
p(A|B)=\hbox{Tr}\, \rho_{AB},\eqno(27)
$$
where
$$
\rho_{AB}=\langle P_A P_B\,\rho P_B P_A\rangle /
\hbox{Tr}\,\langle P_B\,\rho
P_B\rangle\eqno(28)
$$
does obey the constraints.

\hangindent=25pt\hangafter=-1 (b) The conditional probability formula
(8) does
not give the
right propagators. For example, if
$$
\eqalignno{P_{T_1} P_{T_2} &= P_{T_2} P_{T_1} =0,
&(29)\cr\noalign{\vskip3pt\medbreak}
P_{A_1} P_{T_1} &=P_{T_1} P_{A_1}\ne 0,
&(30)\cr\noalign{\vskip3pt\medbreak}
P_{A_2} P_{T_2} &=P_{T_2} P_{A_2} \ne 0, &(31)\cr}
$$
then Kucha\v r wants $p(A_2,T_2\,|\, A_1,T_1)$ to be the absolute
square of the propagator (i.e., the transition probability) to go
from $A_1$ at $T_1$ to $A_2$ at $T_2,$ whereas the formula
$$
p(A_2,T_2\,|\,A_1,T_1)
={\hbox{Tr}\,(P_{A_2}P_{T_2}P_{A_1}P_{T_1}\,\rho\,
P_{T_1}P_{A_1}P_{T_2}P_{A_2})\over
\hbox{Tr}\,(P_{A_1}P_{T_1}\,\rho P_{T_1}P_{A_1})}\eqno(32)
$$
gives zero.

\hangindent=25pt\hangafter=-1 In response, I would say that the
absolute square
of the propagator
is
$$
\big|\langle A_2\hbox{ at }T_2\mid A_1\hbox{ at
}T_1\rangle\big|^2=p(A_2\hbox{ at }T_2\mid A_1\hbox{ at
}T_1),\eqno(33)
$$
that is, the probability to have $A_2$ {\it at} $T_2$ if one had
$A_1$ {\it at} $T_1,$ whereas Eq.~(32) gives
$$
p(A_2,T_2\mid A_1,T_1)=p(A_2\hbox{  and }T_2\mid A_1\hbox{ and
}T_1),\eqno(34)
$$
the probability that one has $A_2$ {\it and} $T_2$ if one has
$A_1$ {\it and} $T_1$ on the same hypersurface. Since Eq.~(29)
says that $T_1$ and $T_2$ are mutually exclusive, this probability is
zero: one cannot simultaneously have two distinct clock readings.

\hangindent=25pt\hangafter=-1 The conditional probability
interpretation is
dynamical in the sense
that it gives the $T$ dependence of $p(A|T)$ by Eq.~(16). However, it
is not so dynamical as Kucha\v r would like to give the transition
probability of Eq.~(33) directly. The reason it does not is that this
quantity is not directly testable, since $T_1$ and $T_2$ are
different conditions that cannot be imposed together. At $T=T_2,$ one
has no direct access to what happened at $T=T_1.$ One must instead
rely upon records, which can be checked at $T_2.$

\hangindent=25pt\hangafter=-1 One way that one could try to calculate
the
transition probability
(33) theoretically is to try to construct a projection operator
$P_{A_1T_1}\ \ (\ne P_{A_1}P_{T_1})$ which, when applied to $\rho,$
gives a density matrix
$$
\rho_{A_1T_1}=P_{A_1T_1}\,\rho
P_{A_1T_1}/\hbox{Tr}\,(P_{A_1T_1}\,\rho)\eqno(35)
$$
which satisfies the constraints and which gives
$$
p(A_1|T_1;\rho_{A_1T_1})
={\hbox{ Tr}\,(P_{T_1}P_{A_1}P_{T_1}\,\rho_{A_1T_1})\over
\hbox{Tr}\,(P_{T_1}\,\rho_{A_1T_1})}=1.\eqno(36)
$$
Then one could say
$$
p(A_2\hbox{ at }T_2\mid A_1\hbox{ at
}T_1)=p(A_2|T_2;\,\rho_{A_1T_1}).\eqno(37)
$$
However, it is not clear whether the answer is unique and whether it
has certain desirable properties. More importantly, it is not
directly testable, and therefore it is rather {\it ad hoc} which
definition to propose for the transition probability.

\section 4. The variation of entropy with clock time.
\hangindent=25pt\hangafter=-1 The time asymmetry of the second law of
thermodynamics should of
course be expressed in terms of physical clock time rather than in
terms of unobservable coordinate time (Page 1992b).

\hangindent=25pt\hangafter=-1 How to express entropy is more
problematic. One
mathematical
expression is
$$
\eqalignno{ S_T &=-\hbox{Tr}\, \rho_T\,\ln\rho_T, &(38)\cr\vvs
\rho_T
&=P_T\,\overline\rho\,P_T/\hbox{Tr}\,(P_T\,\overline\rho\,P_T).
&(39)\cr}
$$
\ \ \ \ \ \ \ \ If
$$
\overline \rho=|\psi\rangle\langle\psi|,\eqno(40)
$$
a pure state, then
$$
\rho_T=|\psi_T\rangle\langle \psi_T|\eqno(41)
$$
with
$$
|\psi_T\rangle= P_T|\psi\rangle /\langle
\psi|P_T|\psi\rangle^{1/2},\eqno(42)
$$
another pure state, so $S_T=0.$

\hangindent=25pt\hangafter=-1 But if $\overline\rho$ is not pure
(e.g., is
obtained from
$$
\rho=|\psi\rangle\langle\psi|\eqno(43)
$$
pure but time dependent), then $S_T$ can exceed zero and vary with
the clock time $T.$

\hangindent=25pt\hangafter=-1 {\sl Example}: Consider two coupled
$\hbox{spin-}{1\over2}$
systems in a vertical
magnetic field, with Hamiltonian
$$\eqalignno
{H &={1\over4} {\vec\sigma}_C\cdot {\vec\sigma}_R+{1\over2}
\sigma_{C_z}\otimes I_R+{1\over2}I_C\otimes
\sigma_{R_z}+{3\over 4} I &\cr\vvs
&=2\,|\uparrow\uparrow\rangle
\langle\uparrow\uparrow|+{1\over2}(|\uparrow
\downarrow\rangle
+|\downarrow\uparrow\rangle)(\langle\uparrow\downarrow|
+\langle\downarrow\uparrow|),
&(44)\cr}
$$
giving a triplet and a singlet degenerate with the lowest state of
the triplet. Take the time-dependent pure state
$$
|\psi\rangle =0.1\sqrt 5\big[(e^{-it}+1)|\uparrow\downarrow\rangle
+(e^{-it}-1)|\downarrow\uparrow\rangle
+4\,|\downarrow\downarrow\rangle\big],\eqno(45)
$$
so the coordinate-time-averaged density matrix is
$$
\eqalignno{\overline\rho &=0.1\big(|\uparrow\downarrow\rangle
\langle\uparrow
\downarrow|+2|\uparrow\downarrow\rangle\langle\downarrow\downarrow|
+2|\downarrow\downarrow\rangle\langle\uparrow\downarrow|
+|\downarrow\uparrow\rangle \langle\downarrow\uparrow| &\cr\vvs
&\qquad\qquad -2|\downarrow\uparrow\rangle
\langle\downarrow\downarrow|
-2|\downarrow\downarrow\rangle\langle\downarrow\uparrow|
+8|\downarrow\downarrow\rangle\langle\downarrow\downarrow|\big).
&(46)}
$$
Let
$$
P_T={1\over 2}\big(|\uparrow\rangle +e^{-iT}\,
|\downarrow\rangle\big)\big(\langle\uparrow|
+e^{iT}\langle\downarrow|\big)_C\otimes I_R.\eqno(48)
$$
This leads to
$$
\rho_T=\rho_{TC}\otimes \rho_{TR}\eqno(49)
$$
with
$$
\eqalignno{\rho_{TC} &={1\over 2}\big(|\uparrow\rangle
+e^{-iT}|\downarrow\rangle\big)\big(\langle\uparrow|
+e^{iT}\langle\downarrow|\big)_C, &(50)\cr\vvs
\rho_{TR} &=(10+4\cos T)^{-1}\big[|\uparrow\rangle\langle\uparrow|
-2|\uparrow\rangle\langle\downarrow|-2|\downarrow\rangle\langle
\uparrow|
+(9+4\cos T)|\downarrow\rangle\langle\downarrow|\big].
&(51)\cr}
$$
Therefore,
$$
\eqalignno{ S(\rho(t)) &=0,\quad S(\overline\rho)=\ln 10-1.8\ln
3\approx
0.3251, &(52)\cr\vvs
S_T=S(\rho_T) &=\ln\left({10+4\cos T\over\sqrt{5+4\cos T}}\right)-
{2\sqrt{1+(2+\cos T)^2}\over 5+2\cos T}&\cr\vvs
&\qquad \times\ln\left({5+2\cos
T+2\sqrt{1+(2+\cos T)^2}\over\sqrt{5+4\cos T}}\right), &(53)\cr}
$$
which varies in a rather complicated way with clock time $T.$

\hangindent=25pt\hangafter=-1 Even when $\rho_T$ is pure and gives
$S_T=0,$ we
may divide the system
into subsystems and add the entropy of each subsystem density matrix
(which ignores the information or negentropy of the correlations
between the subsystems):
$$
\eqalignno{S_{T,
{\rm coarse}} &=
-\sum_{i=1}^n\hbox{tr}\,\rho_{T_i}\,\ln\,\rho_{T_i}, &(54)\cr\vvs
\rho_{T_i} &=
\mathop{\hbox{tr}}_{j\ne i} P_T\,
\overline\rho P_T/\hbox{Tr}\,(P_T\,\overline\rho).
&(55)\cr}
$$

\hangindent=25pt\hangafter=-1 {\sl Example}: Consider three
$\hbox{spin-}{1\over 2}$ systems in a
vertical magnetic field, with Hamiltonian
$$
\eqalignno{H &={1\over 2}\sigma_{C_z}\otimes I_{R1}\otimes I_{R2}
+{1\over 2}I_C\otimes\sigma_{R1_z}\otimes I_{R2}
&\cr\noalign{\vskip3pt\medbreak}
&\qquad \quad +{1\over 2}I_C\otimes
I_{R1}\otimes\sigma_{R2_z}+{1\over 4} I_C\otimes
{\vec\sigma}_{R1}\cdot
{\vec\sigma}_{R2}+{1\over 4}I &\cr\noalign{\vskip3pt\medbreak}
&=2|\uparrow\uparrow\uparrow\rangle\langle\uparrow\uparrow\uparrow|
&\cr\noalign{\vskip3pt\medbreak}
&\qquad \quad +{1\over
2}\big(|\uparrow\uparrow\downarrow\rangle+|\uparrow\downarrow\uparrow
\rangle
\big)\big(\langle\uparrow\uparrow\downarrow|+\langle
\uparrow\downarrow\uparrow|\big)+|\downarrow\uparrow\uparrow\rangle
\langle\downarrow\uparrow\uparrow| &\cr\noalign{\vskip3pt\medbreak}
&\qquad\quad -{1\over 2}\big(|\downarrow\uparrow\downarrow\rangle
-|\downarrow\downarrow\uparrow\rangle\big)
\big(\langle\downarrow\uparrow\downarrow|-\langle\downarrow\downarrow
\uparrow|
\big)-|\downarrow\downarrow\downarrow\rangle
\langle\downarrow\downarrow\downarrow|,
&(56)\cr}
$$
so the first spin acts as a noninteracting clock and the remaining
two have the same spin-spin coupling as the previous example. Take a
pure-state linear combination of the zero-energy eigenvectors,
$$
|\psi\rangle ={1\over 2}\big(|\uparrow\uparrow\downarrow\rangle
-|\uparrow\downarrow\uparrow\rangle
+|\downarrow\uparrow\downarrow\rangle
+|\downarrow\downarrow\uparrow\rangle\big),\eqno(57)
$$
so Eq.~(40) holds. Let $P_T$ be given by Eq.~(48) again, except that
now $I_R$ is the $4\times 4$ unit matrix for the last two spins. This
gives $\rho_T$ of the pure form~(41) with
$$
|\psi_T\rangle =2^{-1/2}\big(|\uparrow\rangle
+e^{-iT}|\downarrow\rangle\big)_C
\otimes {1\over2}\big[(e^{iT}+1)|\uparrow\downarrow\rangle
+(e^{iT}-1)|\downarrow\uparrow\rangle\big]_R,\eqno(58)
$$
so $\rho_T$ again factorizes as in Eq.~(49) between the clock and the
rest, with $\rho_{TC}$ given by Eq.~(50) and
$$
\eqalignno{\rho_{TR} &={1\over 2}\big[(1+\cos
T)|\uparrow\downarrow\rangle
\langle\uparrow\downarrow|-i\sin T|\uparrow\downarrow\rangle
\langle\downarrow\uparrow| &\cr\vvs
&\qquad\qquad +i\sin T |\downarrow\uparrow\rangle
\langle\uparrow\downarrow|+(1-\cos T)|\downarrow\uparrow\rangle
\langle\downarrow\uparrow|\big], &(59)\cr}
$$
a pure state which does  not factorize between the last two spins.
The reduced density matrices of the two spins are
$$
\eqalignno{\rho_{TR1} &={1\over 2}I_{R1}+{1\over 2}\cos T\
\left(|\uparrow\rangle\langle\uparrow|
-|\downarrow\rangle\langle\downarrow|\right)_{R1}, &(60)\cr\vvs
\rho_{TR2} &={1\over 2}I_{R2}-{1\over 2}\cos T
\left(|\uparrow\rangle\langle\uparrow|
-|\downarrow\rangle\langle\downarrow|\right)_{R2}, &(61)\cr}
$$
each with eigenvalues
$$
p_1 =\cos^2{1\over 2}T,\qquad p_2=\sin^2{1\over 2}T.\eqno(62)
$$
Hence, the coarse-grained entropy given by Eq.~(54) is
$$
\eqalignno{S_{T,{\rm coarse}}
&=S(\rho_{TC})+S(\rho_{TR1})+S(\rho_{TR2}) &\cr\vvs
&=-2\cos^2{1\over 2}T\,\ln\cos^2{1\over 2}T-2\sin^2{1\over
2}T\,\ln\sin^2{1\over 2}T, &(63)\cr}
$$
oscillating between $0$ at $T=n\pi$ and $2\ln 2$ at $T=(n+1/2)\pi.$

\hangindent=25pt\hangafter=-1 Although they can vary with clock time,
the
mathematical and
coarse-grained entropies defined by Eqs.~(38) and (54) above are not
really observables. One would like a truly observable entropy
operator $\widehat S$ and then define the clock-time-dependent
entropy as its conditional expectation value:
$$
S_{T,{\rm observable}}=E(\widehat S|T)=\hbox{Tr}\,(P_T\widehat S
P_T\,\overline\rho)/\hbox{Tr}\,(P_T\,\overline\rho).\eqno(64)
$$

\hangindent=25pt\hangafter=-1 Can one find suitable definitions of
$\widehat S$
and $T$ so that
$S_{T,{\rm observable}}$ increases fairly monotonically with $T$ for
the actual density matrix $\rho$ of the universe?

\hangindent=25pt\hangafter=-1 I have benefited from discussions on
these points
at the conference
with Bryce DeWitt, Murray Gell-Mann, James Hartle, Karel Kucha\v r,
Emil Mottola and William Unruh. Financial assistance has been
provided by the Canadian Institute for Advanced Research, the
Fundaci\'on Banco Bilbao Vizcaya, the National Science Foundation,
and the Natural Sciences and Engineering Research Council.

\vskip 1truecm
\centerline{\bf References}

\vskip5truept
Feynman, R.P. and Hibbs, A.R. (1965) {\it Quantum Mechanics and Path
Integrals,} McGraw-Hill, New York.

Gell-Mann, M. and Hartle, J.B. (1990) In {\it Complexity, Entropy,
and the Physics of Information, SFI Studies in the Sciences of
Complexity,} Vol.~VIII, Ed. W.~Zurek, Addison Wesley, Reading, and in
{\it Proceedings of the 3rd International Symposium on the
Foundations of Quantum Mechanics in the Light of New Technology,}
Eds. S.~Kobayashi, H.~Ezawa, Y.~Murayama, and S.~Nomura, Physical
Society of Japan, Tokyo.

Griffiths, R. (1984) {\it J. Stat. Phys.}, {\bf 36}, 219.

Hartle, J.B. (1990a) In {\it Gravitation and Relativity 1989:
Proceedings of the 12th International Conference on General
Relativity and Gravitation,} Eds. N.~Ashby, D.~F. Bartlett, and
W.~Wyss, Cambridge University Press, Cambridge.

Hartle, J.B. (1990b) In {\it Proceedings of the 60th Birthday
Celebration of M.~Gell-Mann,} Eds. J.~Schwarz and F.~Zachariasen,
Cambridge University Press, Cambridge.

Hartle, J.B. (1991) In {\it Quantum Cosmology and Baby Universes,}
Eds. S.~Coleman, J.~B. Hartle, T.~Piran, and S.~Weinberg, World
Scientific, Singapore.

Kucha\v r, K. (1992)
In {\it Proceedings of the 4th Canadian Conference on
General Relativity and Relativistic Astrophysics,} Eds.
G.~Kunstatter, D.~Vincent, and J.~Williams, World Scientific,
Singapore.

Misner, C.W., Thorne, K. S., and Wheeler, J.A. (1973) {\it
Gravitation,} Freeman, San Francisco.

Omn\`es, R. (1988a) {\it J. Stat. Phys.}, {\bf 53}, 893.

Omn\`es, R. (1988b) {\it J. Stat. Phys.}, {\bf 53}, 933.

Omn\`es, R. (1988c) {\it J. Stat. Phys.}, {\bf 53}, 957.

Omn\`es, R. (1989) {\it J. Stat. Phys.}, {\bf 57}, 357.

Page, D.N. (1986a) In {\it Quantum Concepts in Space and Time}, Eds.
R.~Penrose and C.~J. Isham, Clarendon Press, Oxford.

Page, D.N. (1986b) {\it Phys. Rev.}, {\bf D34}, 2267.

Page, D.N. (1987) In {\it String Theory -- Quantum Gravity and
Quantum Cosmology (Integrable and Conformal Invariant Theories):
Proceedings of the Paris-Meudon Colloquium,} Eds. H.~DeVega and
N.~Sanchez, World Scientific, Singapore.

Page, D.N. (1988) In {\it Proceedings of the Fourth Seminar on
Quantum Gravity}, Eds. M.~A. Markov, V.~A. Berezin, and V.~P. Frolov,
World Scientific, Singapore.

Page, D.N. (1989) Time as an inaccessible observable. University of
California at Santa Barbara report NSF-ITP-89-18.

Page, D.N. (1991a) In {\it Conceptual Problems in Quantum Gravity},
Eds. A.~Ashtekar and J.~Stachel, Birkh\"auser, Boston.

Page, D.N. (1991b) In {\it Gravitation: A Banff Summer
Institute,} Eds. R.~Mann and P.~Wesson, World
Scientific, Singapore.

Page, D.N. (1992a) A physical model of the universe. To be published
in {\it The Origin of the Universe: Philosophical and Scientific
Perspectives,} Eds. R.~F. Kitchener and K.~T. Freeman, State
University of New York Press, Albany.

Page, D.N. (1992b) The arrow of time. To be published in {\it
Proceedings of the First International A.D. Sakharov Memorial
Conference on Physics,} Ed. M.~Man'ko, Nova Science Publishers,
Commack.

Page, D.N. and Wootters, W.K. (1983) {\it Phys. Rev.}, {\bf D27},
2885.

Strocchi, F. and Wightman, A.S. (1974) {\it J. Math. Phys.}, {\bf
15}, 2198.

Wheeler, J.A. (1978) The ``past'' and the ``delayed-choice''
double-slit experiment. In {\it Mathematical Foundations of Quantum
Theory,} Ed. A.~R. Marlow, Academic Press, New York.

\vfill\eject
\parindent=25pt
\centerline{\bf Discussion}
\bigskip
Karel Kucha\v r: \quad You always apply the conditional probability
formula
to calculate the conditional probability of a projector at a single
instant of an internal clock time. You never apply it to answering
the
fundamental DYNAMICAL question of the internal Schr\"odinger
interpretation,
namely, ``If one finds the particle at $Q^\prime$ at the time
$T^\prime,$
what is the probability of finding it at $Q^{\prime\prime}$ at the
time
$T^{\prime\prime}?$'' By your formula, that conditional probability
differs
from zero only if $T^\prime=T^{\prime\prime}$
and $Q^\prime=Q^{\prime\prime}.$ In brief, your interpretation
prohibits the
time to flow and the system to move!

For me, this virtually amounts to a reductio ad absurdum of the
conditional
probability proposal. One can trace this feature back to the fact
that the
conventional derivation of the conditional probability formula
amounts
to the violation of the super-Hamiltonian constraint.

Don Page: \quad At this time I cannot give much more of an answer
than
I did in my lecture (which is slightly expanded for the Proceedings).
In my viewpoint, only quantities at a single instant of time are
directly
accessible,
and so one cannot directly test the two-time probability you discuss.
One could instead at one time test the conditional probability that
the particle is at $Q^{\prime\prime},$ given that the time is
$T^{\prime\prime}$
and that at this time there is a {\it record} indicating that the
particle was
at $Q^\prime$ at the time $T^\prime.$ However,  there is no direct
way to test
whether the record is accurate, though one can check whether
different
records show consistency. After all, that is the only way we have to
increase
our confidence in the existence of historical events.

\bigskip
Karel Kucha\v r: \quad
Don, you are the first person I met who simultaneously believes in
the
existence of many worlds and is a solipsist of an instant.

Don Page: \quad I believe that different
instants, i.e., different clock times, are actually
examples of the different worlds. They all exist, but each
observation, and its associated conditional probability, occurs at
one single time (assuming that the condition includes or implies a
precise value of the clock time in question). We can only directly
observe and be aware of the world, and the time, in and at which we
exist, though the correlations in memories and other structures we
observe in one world give indirect evidence of other worlds, and
other clock times, in the full quantum state of the universe.
I do not deny the existence of historical events at different
instants of clock
time, but I do not believe that they have conditional probabilities
(given our conditions here and now) exactly equal to unity, or even
that
they can be precisely assigned any probabilities at all, say in the
sense
of Eq.~(7) when Eq.~(6) holds. In any case, any imputed probabilities
for events in the past or future cannot be directly tested.

\bigskip
\end